\documentstyle[12pt,epsf]{article}

\begin{document}

\newcommand{\sheptitle}
{Charge and Colour Breaking Constraints in the 
MSSM With Non-Universal SUSY Breaking} 

\newcommand{\shepauthor}
{S.~A.~Abel$^a$ and C.~A.~Savoy$^{b}$}

\newcommand{\shepaddress}
{
{\small $^a$Theory Division, Cern 1211, Geneva 23, Switzerland}\\
{\small $^b$CEA-SACLAY, Service de Physique Th\'eorique,}\\
{\small F-91191 Gif-sur-Yvette Cedex, France}    }

\newcommand{\shepabstract}
{We examine charge/colour breaking along 
directions in supersymmetric field space which are $F$ and $D$-flat. 
We catalogue the dangerous directions
and include some new ones which have not previously been 
considered. Analytic expressions for the 
resulting constraints are provided which 
are valid for all patterns of supersymmetry breaking.
As an example we consider a recently proposed pattern 
of supersymmetry breaking derived in Horava-Witten $M$-theory, 
and show that there is no choice of parameters for which the 
physical vacuum is a global minimum.}

\begin{titlepage}
\begin{flushright}
CERN-TH/98-315\\
hep-ph/9809498\\
\end{flushright}
\vspace{0.5in}
\begin{center}
{\large{\bf \sheptitle}}
\bigskip \\ \shepauthor \\ \mbox{} \\ {\it \shepaddress} \\ 
\vspace{0.5in}
{\bf Abstract} \bigskip \end{center} \setcounter{page}{0}
\shepabstract
\vspace{0.5in}
\begin{flushleft}
CERN-TH/98-315\\
\today
\end{flushleft}
\end{titlepage}

%       macros
\newcommand{\hepph}[1]{{\tt hep-ph/#1 }}
\newcommand{\hepth}[1]{{\tt hep-th/#1 }}
\newcommand{\phrd}[3]{{{\it Phys.~Rev.}~{\bf D#1} (#3) #2}}
\newcommand{\plb}[3]{{{\it Phys.~Lett.}~{\bf B#1} (#3) #2}}
\newcommand{\prd}[3]{{{\it Phys.~Rev.}~{\bf D#1} (#3) #2}}
\newcommand{\prl}[3]{{{\it Phys.~Rev.~Lett}~{\bf #1} (#3) #2}}
\newcommand{\npb}[3]{{{\it Nucl.~Phys.}~{\bf B#1} (#3) #2}}
\newcommand{\ptp}[3]{{{\it Prog.~Theor.~Phys.}~{\bf #1} (#3) #2}}
\newcommand{\rpp}[3]{{{\it Rept.~Prog.~Phys.}~{\bf #1} (#3) #2}}
\newcommand{\leqsim}{\,\raisebox{-0.6ex}{$\buildrel < \over \sim$}\,}
\newcommand{\geqsim}{\,\raisebox{-0.6ex}{$\buildrel > \over \sim$}\,}
\newcommand{\be}{\begin{equation}}
\newcommand{\ee}{\end{equation}}
\newcommand{\ba}{\begin{eqnarray}}
\newcommand{\ea}{\end{eqnarray}}
\newcommand{\etal}{\mbox{\em et al}}
\newcommand{\ie}{\mbox{\em i.e.~}}
\newcommand{\eg}{\mbox{\em e.g.~}}
\newcommand{\cf}{\mbox{\em c.f.~}}
\newcommand{\nn}{\nonumber}
\newcommand{\dif}{\mbox{d}}
\def\gev{\,{\rm GeV}}
\def\tev{\,{\rm TeV}}
\newcommand{\smallfrac}[2]{\frac{\mbox{\small #1}}{\mbox{\small #2}}}

\newpage

%%%%%\section{Introduction}

Supersymmetric models often have charge and/or colour breaking minima
which can compete with the physical vacuum~\cite{ccb1,sabel}. 
Excluding the regions of parameter space that lead to such minima 
results in bounds which can be severe\footnote{The alternative, 
to devise a cosmological mechanism to place us 
in the metastable physical vacuum, was discussed 
in Ref.\cite{riotto,us} and will not be considered here. 
In this paper we will be concerned with
the analytic determination of these bounds and shall therefore 
take a pragmatic approach; even if cosmology does place us in the 
physical vacuum we would, at the very least, like to be able to calculate 
if the present vacuum is unstable.}. There are a number of reasons why 
supersymmetry suffers from this kind of metastability.
There are many flat directions which are lifted only 
by our choice of supersymmetry breaking parameters.
The higgs scalar which couples to the bottom quark, 
$H_1$, has the same quantum numbers as the leptons.
Moreover, the mass-squared parameter of the
higgs which couples to the top quark, $H_2$, is negative 
at the weak scale (a requirement of electroweak symmetry breaking and 
one of the successes of supersymmetry). 

The resulting charge and colour breaking bounds can be roughly 
divided into three kinds. 
\begin{itemize}
\item $D$-flat directions which develop a minimum due to 
       large trilinear supersymmetry breaking terms. They are important 
       at low scales~\cite{ccb1} and are usually referred to as CCB bounds. 
\item $F$ and $D$ flat directions corresponding to a single gauge invariant.
These are important when there 
are negative mass-squared terms at the GUT scale. One therefore needs 
to take account of the renormalisation group running, but they can be 
lifted by non-renormalisable terms in the superpotential. 
They are often referred to as 
unbounded from below (UFB) bounds~\cite{riotto}.
\item $F$ and $D$ flat directions which correspond to a combination 
of gauge invariants involving $H_2$. 
These can develop a minimum at intermediate scales due to the running 
$H_2$ mass even if all the mass-squareds are positive at the GUT 
scale. They {\em cannot} be lifted by non-renormalisable terms.
They are also (rather confusingly) referred to as UFB bounds.
\end{itemize}

The 3rd kind of bound will be of interest here since
they can be very severe. 
For example, the Constrained MSSM (CMSSM) 
near its low $\tan\beta $ fixed point has almost half its parameter space 
excluded by them. Combined with dark matter constraints and 
experimental bounds, this is sufficient to exclude the model 
entirely~\cite{sabel}.
Unfortunately these bounds are also the most difficult to calculate 
because the renormalisation group running of the soft 
supersymmetry breaking parameters plays a complicated role
in their determination. Specifically, the dangerous minimum 
forms where the running (negative) $H_2$ mass-squared is 
offset by the other (positive) mass-squared terms.  

Hence most work on UFB bounds is done 
numerically and the models considered typically 
have few parameters (such as the CMSSM which has only four).
The reason for this is a practical one; to properly sample a parameter 
space one has to include, say, one low and one high
representative of each variable. This becomes progressively more 
difficult as the dimensionality increases and for the most general 
MSSM (which has $\sim 100$ parameters) it is clearly impossible. 
Even if it were possible to sample such a large parameter space, 
extracting any useful information would be extremely difficult.

However, what does one do if a  
case of phenomenological interest (the real World for example)
requires a less restricted set of parameters than that of the CMSSM?
How do the UFB bounds restrict supersymmetry breaking in more 
general models? While determining this would very difficult numerically,
it was shown in Ref.\cite{us} that, by making a few simple approximations, 
we can make a reasonably accurate analytic approximation. In this 
letter we shall present the analytic expressions 
for the UFB bounds in the most general supersymmetry breaking scenarios. 
In particular we will show that there is essentially
only one particular combination of supersymmetry breaking parameters which 
is subject to UFB bounds. This combination is 
\be
\label{combi}
2 m^2_{L_{ii}} +  m_2^2
- m_{U_{33}}^2 - m_{Q_{33}}^2,
\ee
where the notation for soft supersymmetry breaking parameters is as in 
Ref.\cite{us}. 

We begin by ennumerating the dangerous directions,
assuming the usual $R$-parity invariant superpotential of 
the MSSM; 
\be
W_{MSSM}=h_U Q H_2 U^c + h_D Q H_1 D^c + h_E L H_1 E^c+\mu H_1 H_2.
\ee
As we stated above, the dangerous $F$ and $D$ flat directions are those 
constructed from gauge invariants involving $H_2$~\cite{carlos,tony}. 
The first example of this kind in the literature is (see Komatsu 
in Ref.\cite{ccb1})
\be
L_i Q_3 D_3 \mbox{ ; } H_2 L_i
\ee 
where the suffices on matter superfields are generation indices. 
With the following choice of VEVs;
\ba
\label{komkom}
h_2^0             &=& -a^2 \mu/h_{D_{33}} \nonumber \\
\tilde{d}_{L_3}=\tilde{d}_{R_3} &=& a \mu/h_{D_{33}} \nonumber \\
\tilde{\nu}_i   &=& a  \sqrt{1+a^2} \mu/h_{D_{33}}, 
\ea
the potential along this direction is $F$ and $D$-flat, and
depends only on the soft supersymmetry breaking terms;
\be
\label{softv}
V=\frac{\mu ^2}{h_{D_{33}}^2} a^2 (a^2 (m_2^2+m_{L_{ii}}^2) + 
m_{L_{ii}}^2+m_{d_{33}}^2+m_{Q_{33}}^2 ).
\ee 
This is not quite, but is very close to, the deepest `fully optimised'
direction~\cite{ccb1}. 
Two additional points. Note that in this case we can work in 
the basis in which the relevant Yukawa coupling is diagonal. Also, 
since the only severe bounds involve the third generation we shall 
restrict our attention to these cases. 

At large values of $a\gg 1$ the potential is governed by the 
first term. This can be negative at the weak scale because 
of the negative value of $m_2^2$. 
The potential can therefore develop a charge and colour breaking 
minimum at a scale of $few\times \mu /h_{D_j} $ and ensuring that this does 
not happen leads to the UFB bound\footnote{As discussed in Ref.\cite{us}, 
the traditional bound -- that the physical vacuum is the global minimum --
is very close to a sufficient condition -- that the physical vacuum 
be the {\em only} minimum. The latter condition circumvents any cosmological 
questions concerning, for example, the tunneling rate between minima.}.

The coefficients in Eq.(\ref{komkom}) were chosen to cancel 
$F_{H_1}=\partial W/\partial {H_1}$. However there are other 
combinations of invariants which can do the same job; in fact 
$L_k H_2$ plus any invariant containing either $L_3 E_3 $ or 
$Q_3 D_3$ can cancel $F_{H_1}$. However the invariant should not contain 
$H_1$ otherwise $F_E$, $F_D$, $F_Q$ and $F_L$ would all be non-zero. The complete 
basis of gauge invariant monomials in Ref.\cite{tony} then tells us that
the following invariants are of possible relevance;
\ba
\label{invs}
& L_iL_3E_3 & i\neq 3 \nn\\
& L_iQ_3D_3 & \nn\\
& Q'_iU_jL_3E_3  & \nn\\
& Q'_iU_jQ'_3D_3 & i\neq 3  \nn\\
& (Q'_1L_i)(Q'_2L_j)(Q'_3L_3)E_3& \mbox{\hspace{-0.2cm} $+ L$ perms}   \nn\\
& (U_1U_2D_i) Q'_3D_j H_2 & \delta_{3i}+\delta_{3j} \neq 0\nn\\
& (U_1U_2D_i) Q'_3D_j Q'_lD_m & l\neq 3; \delta_{3i}+\delta_{3j}+
\delta_{m3} \neq 0,
\ea
where groups of three coloured fields are contracted with Levi-Cevita symbols.
Note that in the above list, $Q'$ is in the up-quark mass basis to ensure 
that $F_{H_2}=0$. 
The first two directions have already been considered at some length 
in the literature. However this will be, to our knowledge, the first time 
that the other directions have been discussed. 

We can discard most of the new directions because they have non-zero 
$F_Q$ or $F_U$. For example both the $UUDQDH_2$ and $UUDQDQD$ directions have
non-zero $F_{Q'_2}$. This is enough to lift the radiative minima 
since the latter has a depth of $\sim  a_{min}^4 \mu^2 m_W^2/h_b^2$ 
where $a_{min}={\cal O}(10)$, but 
\be
F_{Q'_2}^2 = h_c^2 a_{min}^6 \mu^4 /h_b^4
\gg a_{min}^4 \mu^2 m_W^2/h_b^2 .
\ee
Clearly the only dangerous directions involve only the first generation 
up-quarks. 
For example, the $Q'_1U_1L_3E_3$ direction {\em is} dangerous. 
Although it has non-zero $F_{U_1}$, 
$F_{Q'_1}$ and $F_{H_2}$, these $F$-terms are not enough to 
lift the radiatively induced minima since 
\be
a_{min}^2 F^2_{H_2}=F_{U_1}^2 = F_{Q'_1}^2 
\approx  h_u^2 a_{min}^6 \mu^4 /h_b^4
\ll a_{min}^4 \mu^2 m_W^2/h_b^2 .
\ee
Similar consideration apply for the $Q'UQ'D$ direction. Finally 
the $Q'Q'Q'LLLE$ direction has nothing apart from non-zero  
$F_{U_1}$-terms if we choose VEVs along the $h_2^0 \nu_k$ and 
$U'_{L_1}D'_{L_2}D'_{L_3}E_{L_3}\nu_i\nu_j E_3 $ directions.

The potentials for these operators (when combined with $ h^0_2\nu_k$ 
and neglecting the small $F^2_{U_1}$, 
$F^2_{Q'_1}$ and $F^2_{H_2}$ terms above) are
\ba
LLE &\rightarrow &\frac{\mu ^2}{h_{E_{33}}^2} a^2 (a^2 (m_2^2+m_{L_{kk}}^2) + 
\sqrt{a^2+1}(1-\delta_{ik})(m^2_{L_{ik}}+m^2_{L_{ki}}) \nn\\ 
&&
\mbox{\hspace{1cm}}+m_{L_{ii}}^2+m_{L_{33}}^2+m_{E_{33}}^2 ) \nn\\
LQD &\rightarrow &\frac{\mu ^2}{h_{D_{33}}^2} a^2 (a^2 (m_2^2+m_{L_{kk}}^2) +
\sqrt{a^2+1}(1-\delta_{ik})(m^2_{L_{ik}}+m^2_{L_{ki}}) \nn \\
&& \mbox{\hspace{1cm}}+m_{L_{ii}}^2+m_{Q_{33}}^2+m_{D_{33}}^2 ) \nn\\
QULE &\rightarrow &\frac{\mu ^2}{h_{E_{33}}^2} a^2 (a^2 (m_2^2+m_{L_{kk}}^2) + 
m_{Q'_{11}}^2+m_{U_{11}}^2+m_{L_{33}}^2+m_{E_{33}}^2 ) \nn\\
QUQD &\rightarrow &\frac{\mu ^2}{h_{D_{33}}^2} a^2 (a^2 (m_2^2+m_{L_{kk}}^2) + 
m_{Q'_{11}}^2+m_{U_{11}}^2+m_{Q'_{33}}^2+m_{D_{33}}^2 ) \nn\\
QQQLLLE &\rightarrow 
&\frac{\mu ^2}{h_{E_{33}}^2} 
a^2 (a^2 (m_2^2+m_{L_{kk}}^2)+(1-\delta_{ij})(m_{L_{ij}}^2+m_{L_{ji}}^2)  \nn\\
&& \mbox{\hspace{0.2cm}} 
+\sqrt{a^2+1}(1-\delta_{ij})(m^2_{L_{ij}}+m^2_{L_{ji}})+ 
\sqrt{a^2+1}(1-\delta_{ik})(m^2_{L_{ik}}+m^2_{L_{ki}}) \nn\\ 
&&
+m_{Q_{11}}^2+m_{Q_{22}}^2+m_{Q_{33}}^2+m_{Q_{23}}^2+m_{Q_{32}}^2
+m_{L_{ii}}^2+m_{L_{jj}}^2
+ m_{L_{33}}^2 +m_{E_{33}}^2
 ) 
\ea
where the $h_{D_{33}}$ should be the Yukawa coupling taken in
the up-quark diagonal basis for the $QUQD$ direction.
At first glance the last direction seem
to be safe because of the large number of mass-squared terms. 
However, it should be stressed that these parameters
are not physical mass-squareds and they can therefore be negative 
(as is often the case in weakly coupled string models 
for example). As we shall see however, the renormalisation 
group running {\em does} tend to make them positive towards low scales
so that they indeed tend to be less important. 
The off-diagonal bounds are likely to be unimportant when 
the relevant mass-squared parameters are positive, so henceforth 
we shall consider only the diagonal ones. 

In order to obtain the UFB bound we now need to take account of  
the renormalisation group running of the mass-squared parameters 
between the weak and GUT scales. An accurate analytic method
was introduced in Ref.\cite{us} which we shall briefly recap. 

First we need to make some approximations and assumptions; 
they are 
\begin{itemize}
\item Neglect two loop effects in the renormalization 
group running of parameters (here we include hypercharge contributions).
\item Of the Yukawas, keep only $h_t=h_{U_{33}}$ in the runnning -- neglect
the bottom and tau Yukawas (\ie $\tan\beta <30 $) and mixing.
\item Assume that the gaugino masses are degenerate 
($M_3=M_2=M_1=M_{1/2}$) at the scale where the gauge couplings 
unify (which we shall refer to as the GUT scale)
\end{itemize}
These approximations allow us to obtain analytic solutions for 
the running parameters in closed form (which were presented in Ref.\cite{us}).
In Ref.\cite{us} it was shown that the final analytic expressions 
for the UFB bounds should be accurate to $\sim 15 \%$.

Next, we assume that 
the largest mass, and therefore the appropriate scale to evaluate the 
parameters at is $\phi=h_{U_{33}} \langle h_2^0 \rangle $. In 
the above potentials, $ \langle h_2^0 \rangle= -a^2 \mu /
h_{D_{33},E_{33}} $ so that the potentials are of the form 
\be
V=\frac{M_{GUT}^2}{h_{U_{33}}^2} \hat\phi
 \left( \hat\phi A + B/b \right)
\ee 
where $A=m_2^2(\phi)+m_{L_{ii}}^2(\phi)$, $B$ is the other combination 
of mass-squared parameters (also evaluated at $\phi$) which appears 
in the potentials above,
\be 
\hat\phi=\phi/M_{GUT}
\ee
and 
\be
\label{b}
b(\phi) = \frac{M_{GUT}h_{D_{33}}}{h_{U_{33}} \mu}= 
\frac{2.9\times 10^{12}}{ h_{U_{33}}/1.1} 
\left(
\frac{M_{GUT}}{2\times 10^{16} \gev}
\right)
\left(
\frac{m_b(\phi)}{2.5\gev }
\right)
\left(
\frac{200 \gev}{\mu(\phi)}
\right)
\ee
for $h_{D_{33}} $ or
\be
\label{b2}
b(\phi) = \frac{M_{GUT}h_{E_{33}}}{h_{U_{33}} \mu}= 
\frac{2.0\times 10^{12}}{ h_{U_{33}}/1.1} 
\left(
\frac{M_{GUT}}{2\times 10^{16} \gev}
\right)
\left(
\frac{m_\tau(\phi)}{1.78\gev }
\right)
\left(
\frac{200 \gev}{\mu(\phi)}
\right)
\ee
for $h_{E_{33}}$. The traditional UFB bound is saturated by $V=V'=0 $; 
the non-trivial solution is therefore 
also a solution to $\tilde{V}=\tilde{V}'=0$ where 
\be
\tilde{V}=
\hat\phi A + B/b.
\ee 
Hence, only the parameter $b$ enters the bound. In fact the bound always 
becomes {\em more} restrictive with increase in $b$, since this decreases the 
positive contribution to the potential. (The increase with unification 
scale has already been noted in numerical work, \eg Casas \etal~ of 
Ref.\cite{sabel}.) To solve for the UFB bound we first define the point 
$\hat\phi_p$ where 
\be
\label{zero}
A(\phi_p)=0.
\ee
We can then expand about $\phi_p$ and solve 
$\tilde{V}(\phi_d)={\tilde{V}'}(\phi_d)=0$;
\ba
\label{rpd}
\hat{\phi}^2_p &=& \left. \frac{e B}{A'b}\right|_{\phi_p} \nn\\
\phi_d &=& e^{-1}\phi_p.
\ea
Likewise, the sufficient condition is given by, 
(${\tilde{V}''}(\phi_c)=V'(\phi_c)=0$);
\ba
\label{rpc}
\hat{\phi}^2_p &=& \left.\frac{e^{3/2} B}{2 A' b}\right|_{\phi_p} \nn\\
\phi_c &=& e^{-3/2}\phi_p.
\ea
The solutions for the running parameters including $b$ and $A'$ 
were given in Ref.\cite{us} in terms of 
the three parameters with quasi-fixed points;
\ba
R &\equiv & h_{U_{33}}^2/g_3^2 \nonumber \\
A_t &\equiv & A_{U_{33}} \nonumber \\
3 M^2 & \equiv & m_{U_{33}}^2 + 
m_{Q_{33}}^2 +m_{2}^2 .
\ea
Using these solutions we can find the value of 
$\hat{\phi}_p $  by solving Eq.(\ref{rpd}) or Eq.(\ref{rpc}) 
either numerically or iteratively. Finally we obtain the bound 
by solving Eq.(\ref{zero}) (which we haven't yet used);
\be
\label{arp}
A(\phi_p)=0=\smallfrac{3}{2} M^2+\left.(m_2^2+m_{L_{ii}}^2-\frac{3}{2}M^2)
\right|_0
-M_{1/2}^2  (
\smallfrac{8}{9}\delta_3^{(2)}+
\smallfrac{3}{2}\delta_2^{(2)}+
\smallfrac{5}{198}\delta_1^{(2)}
),
\ee
where 
\ba
\tilde{\alpha}_i &=& \frac{\alpha_{i}}{\alpha_i|_0}\nn\\
\delta_i^{(n)} & = & \tilde{\alpha}^n_i-1
\ea
and the $0$-subscript indicates values at the GUT scale. 
Using Eqs.(\ref{rpd},\ref{arp}) we are now able to estimate the UFB 
bounds in any particular case of interest and to get a reasonable 
estimate for the most general case. 

\subsubsection*{General models at the low $\tan\beta $ fixed point}

First consider the low 
$\tan\beta $ fixed point (where $h_{U_{33}}$ is very large at 
the GUT scale). As a rule of thumb the UFB constraints are most 
restrictive here. This is because the dangerous minimum is 
driven by the negative mass-squared of $H_2$ which is in turn 
driven by the top quark Yukawa coupling. A larger top 
Yukawa makes $m_2^2$ negative closer to the GUT scale and 
the other mass-squared terms have to be larger to overcome it;
and the largest possible top Yukawa is at the fixed point. 

Here we see why the bounds involve the particular combination 
appearing in Eq.(\ref{combi}). From Eq.(\ref{arp}) the 
relevant bound is
\be
\label{arp2}
\left.(m_2^2+m_{L_{ii}}^2-\frac{3}{2}M^2)\right|_0
\geqsim 
\smallfrac{3}{2} M^2
-M_{1/2}^2  (
\smallfrac{8}{9}\delta_3^{(2)}+
\smallfrac{3}{2}\delta_2^{(2)}+
\smallfrac{5}{198}\delta_1^{(2)}
).
\ee
The $M^2$ parameter is determined at any particular scale since we are 
at the fixed point. The only influence the $B/b$ term can 
have is logarithmic, via the determination of the scale, 
$\phi_p$, at which the RHS should be evaluated. Consider the $LLE$,~$LH_2$ 
directions with the central value of $b$. Using the solutions  of
Ref.\cite{us} this scale is given by 
\be
\hat\phi_p=\frac{2\pi e}{3 b \alpha |_0 }
\left(\frac{(\tilde{m}_{L_{ii}}^2+\tilde{m}_{L_{33}}^2+
\tilde{m}_{E_{33}}^2)|_0 -3 \delta_2^{(2)}}
{\tilde{\alpha}_3 R ( 3 \tilde{M}^2 + 
\tilde{A}_{U_{33}}^2 )-2 \tilde{\alpha}_2^3}\right)
\ee 
where $\tilde{m}=m/M_{1/2}$, which can be solved iteratively. 
The final bound is 
\be
\label{arp3}
\left.(2 \tilde{m}_{L_{ii}}^2 + \tilde{m}_2^2-\tilde{m}^2_{U_{33}}-
\tilde{m}^2_{Q_{33}})\right|_0
\geqsim 
f(\tilde{B}|_0)
\ee
where 
\ba
\tilde{B}|_0&=&\left.(\tilde{m}_{L_{ii}}^2+\tilde{m}_{L_{33}}^2+
\tilde{m}_{E_{33}}^2)\right|_0\nn\\
f(x)&\approx& 1.43-0.16 x +0.02 x^2.
\ea
The last equation is a fit to the bounds which is accurate for 
$x\geqsim 0$.
We stress that this result is valid for the MSSM with 
{\em any} pattern of supersymmetry breaking. In the CMSSM, where 
the scalar masses are degenerate and equal to $m_0$ 
at the GUT scale, the inequality becomes 
\be 
\tilde{m}_0^2 \geqsim f(3\tilde{m}_0) 
\ee
and we find the familiar result 
\be 
m_0 \geqsim 1.05 M_{1/2}
\ee
which is a slight overestimate of the numerical results of 
Ref.\cite{sabel} (by $ 5-10 \%$). 

We can also examine the dependence on the $b $ parameter 
for this case. For example, the central value of $b$ had $\mu = 200 \gev$. 
If we take $\mu = 500 \gev $ (or $b=7.7\times 10^{11}$) 
we find 
\be
f(x)\approx 1.20-0.14 x +0.02 x^2
\ee
and for the CMSSM we then find
\be 
m_0 \geqsim 0.95 M_{1/2}.
\ee
From this we can conclude that, given the accuracy of our other 
approximations, there is little point in determining the value of 
$\mu $ by minimising the potential (which is fortunate since this 
would have made our task prohibitively difficult). Additionally 
$M_{GUT}$ is as important as $\mu $ in the determination of the bound
and we can reasonably set $\mu = M_{1/2}$ for an estimate of the UFB bound.

Before leaving the fixed point, there is an additional 
fairly weak bound on $B|_0 $. When this parameter becomes 
too negative it is unable to lift the minimum for any GUT scale 
value of $m^2_{L_{ii}}+m_2^2 $. The resulting bound is 
\be
B|_0=\left.(m_{L_{ii}}^2+m_{L_{33}}^2+
m_{E_{33}}^2)\right|_0 > -0.7 M_{1/2}^2 .
\ee

The values of the functions, $f(x)$, and the above bounds are 
given in the table for all the directions in Eq.(\ref{invs}).
As can be seen the constraints corresponding to the $LLE$,~$LH_2$ 
invariants are the strongest. This had been noted before in numerical 
work on the CMSSM, but here we see it is a completely general result. 
\begin{table}[h]
\label{table1}
\hspace{1cm}
\begin{tabular}{|c|c|c|c|}
\hline 
$LH_2$+Operator & $f(x)$ & $g(x)$ & $\tilde{B}_0\geqsim$ \\
\hline
\hline
$LLE$ & $1.43-0.16x+0.02x^2$ & $2.94-0.2x+0.02x^2$ & -0.7 \\
$QDE$ & $1.08-0.05x$  & $2.55-0.06x$ & -3.5 \\
$QULE$ & $1.03-0.04x$ & $2.49-0.05x$ & -4.5 \\
$QUQD$ & $0.94-0.02x$ & $2.38-0.03x$ & -7.2 \\
$QQQLLLE$ & $0.96-0.02x $ & $2.39-0.03x$ & -6.8 \\
%$UUDQDH_2$ & $0.92-0.02x $ & $2.34-0.03x$ & -8.5\\
\hline
\end{tabular}
\end{table}

\subsubsection*{General models away from the fixed point}

At the fixed point, the bounds do not depend on $M^2|_0 $ or 
$A_{U_{33}}|_0\equiv A_0$. As we move away from the fixed point (to higher 
$\tan\beta $) we find that the dependence on these parameters 
increases. We now include this dependence. The distance from the 
fixed point is most conveniently expressed in terms of the parameter 
\be
\rho = \frac{R}{R^{QFP}} 
\ee
where $R^{QFP}$ is the fixed point value of $R$ which is simply a 
function of scale. 

The bounds can be evaluated as above from Eq.(\ref{arp}) 
keeping the dependence on $A_0$ and $M^2|_0$. In order to determine the 
scale $\phi_p$ we can approximate $A'(\phi)$ with its fixed point value which 
is independent of $A_0$ and $M^2|_0 $; the actual value is close to the 
fixed point value even 
for quite large values of $\tan\beta $ and in any case
the scale appears only logarithmically in the determination 
of the UFB bound. The bounds are found to be 
\be
\label{arp4}
\left.(2 \tilde{m}_{L_{ii}}^2 + \tilde{m}_2^2-\tilde{m}^2_{U_{33}}-
\tilde{m}^2_{Q_{33}})\right|_0
\geqsim 
f(\tilde{B}|_0)+(\rho_p-1)\left( g(\tilde{B}|_0)+3 \tilde{M}^2|_0
-\rho_p (1-\tilde{A}_0)^2 \right),
\ee
where $\rho_p $ is the value of $\rho $ at the scale $\phi_p$. 
To a very good approximation we find that the 
value of $\rho_p$ is given by~\cite{us}
\be 
\frac{1}{\rho_p} = 
1+\frac{1}{2 R_0} = 1+3.17 (\sin^2\beta -\sin^2 \beta^{QFP}),
\ee
where $R_0$ is the value of $R$ at the GUT scale and 
$\tan\beta^{QFP}=1.78 $ in this approximation (but it should be stressed 
that the bounds we are evaluating 
are not dependent on its absolute value but rather the value of 
$\rho_p$). In Eq.(\ref{arp4}), $f(x)$ is the bound at the fixed point 
as before, and $g(x) $ can be approximated in a similar manner. 
For the $LLE$,~$LH_2$ direction we find 
\be
g(x)\approx 2.94 -0.20 x + 0.02 x^2 .
\ee
The functions for the remaining directions are given in the table above.
We can see that $A_0 \approx M_{1/2}$ results in the weakest 
bounds. In addition the bounds asymptote at large $\tan\beta $ to 
the values with 
\be
\rho_p = \frac{1}{1+3.17\cos^2 \beta^{QFP}}\approx 0.57.
\ee
To get some estimate of the errors involved in neglecting the
two loop and threshold effects, when they are included in the running
$\tan\beta^{QFP}\approx 1.6$, which would give
\be
\rho_p \approx 0.53.
\ee 

\subsubsection*{Application: UFB bounds in M-theory}

Now let us apply this method to supersymmetry breaking by 
bulk moduli superfields in Horava-Witten $M$-theory~\cite{hw}. 
The supersymmetry breaking terms were derived in Ref.\cite{munoz}, 
for the case where the dilaton, $S$, plus a single modulus field, $T$, 
are responsible for supersymmetry breaking (see Refs.\cite{susyinM} for 
more on this and other methods of supersymmetry breaking in $M$-theory). 
The supersymmetry breaking can be parameterized by defining 
the $F$-terms as
\ba
F^S&=& \sqrt{3} m_{3/2} (S+\overline{S})\sin\theta\nn\\
F^T&=& \sqrt{3} m_{3/2} (T+\overline{T})\cos\theta,
\ea
where $m_{3/2}$ is the gravitino mass and we have (as in Ref.\cite{munoz})
set the phases and the tree-level vacuum energy density to zero. 
The structure of the resulting supersymmetry breaking is substantially 
different from the weakly coupled string, and depends on 
\be
y=\alpha(T+\overline{T})/(S+\overline{S})
\ee
where $\alpha(T+\overline{T}) $ comes from a correction to the 
gauge kinetic function;
\ba
\frac{M_{1/2}}{m_{3/2}}&=& \frac{\sqrt{3}}{1+y}\left(
\sin\theta+\frac{y}{\sqrt{3}}\cos\theta \right)\nn\\
\frac{m^2_0}{m^2_{3/2}}&=&1
- \frac{3}{3+y}
\left(
y\left(2-\frac{y}{3+y}\right)\sin^2\theta+
\left(2-\frac{3}{3+y}\right)\cos^2\theta-
\frac{2\sqrt{3}y}{3+y}\sin\theta\cos\theta\right)\nn\\
\frac{A_0}{m_{3/2}}&=&\sqrt{3}
\left(
\left(-1+\frac{3y}{3+y}\right)\sin\theta+
\sqrt{3}\left(-1+\frac{3}{3+y}\right)\cos\theta
\right).
\ea
Now from Eq.(\ref{arp4}) we can see that in order to 
avoid new charge a colour breaking minima we should satisfy
the inequality 
\be
F(y,\theta)= \tilde{m}^2_0 -f(\tilde{m}^2_0)
+(1-\rho_p)\left( 
g(3 \tilde{m}^2_0)+ 3 \tilde{m}^2_0-\rho_p (1-\tilde{A}_0)^2\right)
\geqsim 0.
\ee
The inequality is most easily satisfied away from the 
fixed point so we choose the asymptotic low value of 
$\rho_p=0.53 $. Fig.(1) shows the resulting contour 
in the ($y,\theta $) plane. At first sight this would seem to be a good result 
since there are some regions where the inequality is satisifed. 
However, in and between the 
two regions our approximations break down because the mass-squared 
parameters are negative at the GUT scale and the potential 
{\em never becomes positive} -- \ie there is no $\rho_p$. 
Between the contours 
the potential is {\em really} unbounded from below with no minimum. The
UFB bound is actually closest to being satisfied at 
$\theta\approx 0.4$, however no reasonable choice of parameters 
($\mu$ or $\tan\beta^{QFP}$ for example) can lift the charge and 
colour breaking minimum.

Thus we conclude that in this model there is no choice of parameters 
for which the physical vacuum is the global minimum. This could have 
been guessed from the fact that (as stated in Ref\cite{munoz}) the 
scalar mass parameters are always less than the gaugino masses. However
showing it numerically would have been difficult. We should also add that 
the minima are only a problem when $R$-parity is conserved. 
$R$-parity violating models with this pattern of supersymmetry breaking 
are safe from UFB constraints provided the $R$-parity violation is 
strong enough and in the correct Yukawa couplings~\cite{us}. 

\subsection*{Acknowledgements}
We would like to thank Toby Falk and Leszek Roszkowski 
for discussions.

\newpage 

\begin{figure}
\label{fig1}
\vspace*{-4in}
\hspace*{-2.0in}
\epsfysize=12in
\epsffile{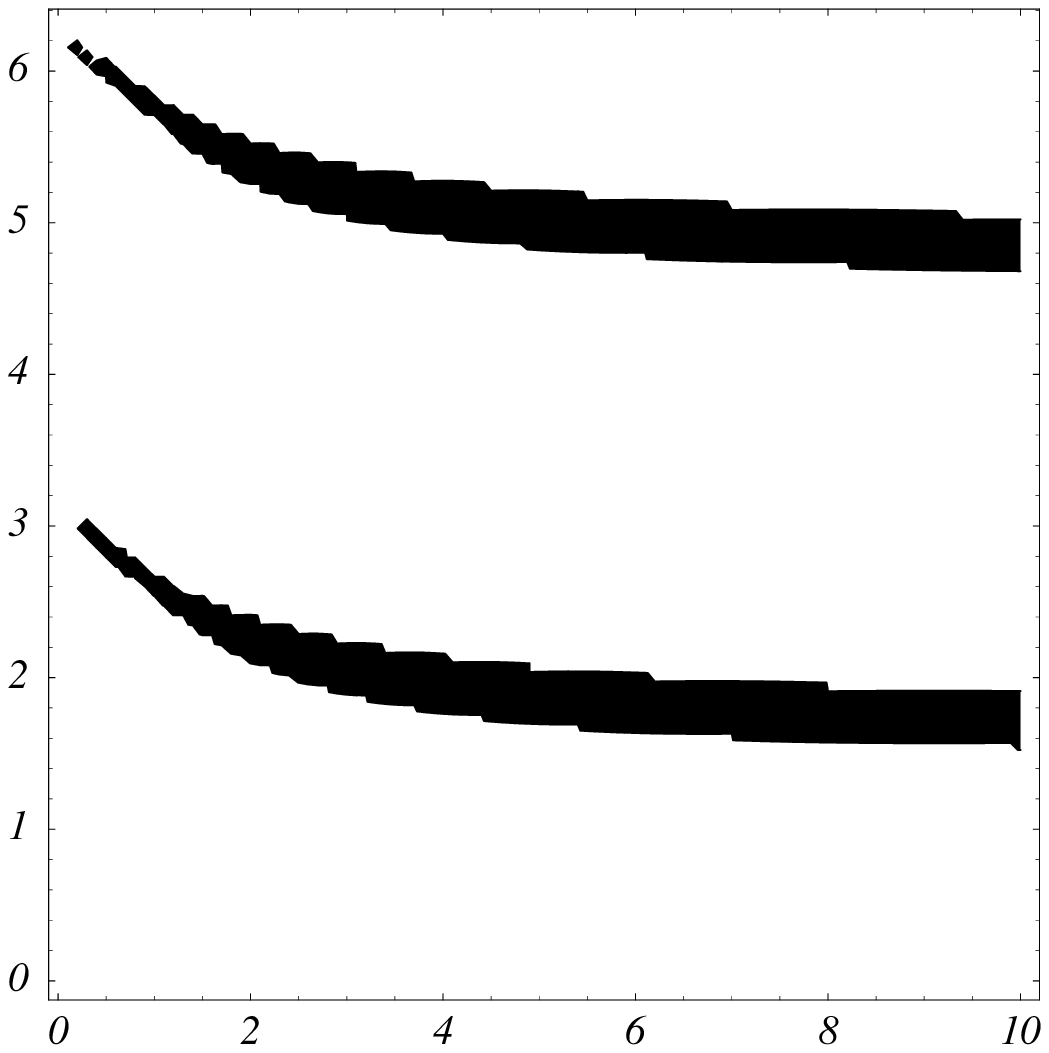}
\vspace{-2in}
\caption{The $ F(y,\theta) =0 $ contour in the $(y,\theta)$ plane.} 
\vspace*{2in}
\end{figure}
\vspace{2in}

\end{document}